\documentclass[journal]{IEEEtran}

\ifCLASSINFOpdf
\else
   \usepackage[dvips]{graphicx}
\fi
\usepackage{url}
\usepackage{hyperref}

\usepackage{graphicx}
\usepackage{amsmath}
\usepackage{multirow}
\usepackage{booktabs}
\usepackage{tabularx}
\usepackage{array}

\begin{document}

\title{Interpretable Singing Emotion Recognition Using Acoustic and Vocal Technique Features}

\author{Fotini Kiriakopoulou, 
        Vassilis Lyberatos, 
        %\IEEEmembership{Member, IEEE},
        Spyridon Kantarelis, Athanasios Voulodimos and Giorgos Stamou
\thanks{Fotini Kiriakopoulou, Vassilis Lyberatos, Spyridon Kantarelis, Athanasios Voulodimos, and Giorgos Stamou are with the National Technical University of Athens, Athens, Greece
(e-mail: fotinikrp@hotmail.com).\par
This work has been submitted to the IEEE for possible publication. Copyright may be transferred without notice, after which this version may no longer be accessible.}
}
% \markboth{IEEE Signal Processing Letters}%
\markboth{}%
{Kiriakopoulou \MakeLowercase{\textit{et al.}}: Interpretable Singing Emotion Recognition}
\maketitle

\begin{abstract}
Singing emotion recognition relies predominantly on deep learning models that achieve strong performance but offer limited insight into the vocal characteristics driving emotion, often overlooking the singing voice as a distinct signal. We propose an interpretable singing emotion recognition framework using a compact set of singing-specific descriptors across prosody, spectral features, phonetics, and vocal technique. To decode model decisions and link predictions to measurable acoustic phenomena, we employ \textsc{SHAP} and apply \textsc{Grad-CAM} to qualitatively assess the learned representations of vocal techniques. Evaluations show that singing-specific features reduce cross-singer variability, while vocal technique descriptors provide interpretable insights into singing expression. Furthermore, out-of-domain evaluation on \textsl{VocalSet} reveals interesting associations between emotional expression and ground-truth technical execution.  The analysis highlights vibrato and pharyngeal techniques as salient cues of emotional delivery, with \textsc{SHAP} indicating that vibrato is context-dependent on articulation rate. These findings emphasize the value of domain-specific features for interpretable singing emotion analysis.
\end{abstract}


\IEEEpeerreviewmaketitle

\section{Introduction}

Singing emotion recognition (SiER) has  relied on deep learning models, which learn complex acoustic representations directly from audio and achieve strong recognition performance~\cite{2024_ravdess, 2025_ravdess}. However, these representations are difficult to relate to specific characteristics of vocal expression, limiting the interpretability of the recognition process. Moreover, their reliance on controlled singing data may not reflect naturally occurring expressive variation. This is particularly relevant in music applications, where understanding the acoustic cues associated with an emotional prediction can be important for analyzing vocal expression, music recommendation, and performance analysis. Explainable AI therefore provides a means of relating model decisions to human-interpretable properties of the input signal~\cite{guidotti2018survey}.

In music emotion recognition, interpretable approaches commonly extract perceptually meaningful acoustic or musical features and use them with relatively shallow classifiers~\cite{xai_music_emotion, lyberatos2024perceptual, patakis25_interspeech}. However, existing analyses of singing rarely incorporate singing-specific descriptors that distinguish singing from speech or instrumental sound.

Singing exhibits acoustic properties that distinguish it from speech, including a wider pitch range, sustained harmonic structure, and distinctive resonance characteristics~\cite{acoustics_sund}. Moreover, emotional expression in singing is closely related to vocal technique. For example, vocal resonance has been shown to influence emotional perception causally~\cite{resonance_emotion}, while singers may use pitch modulation, such as vibrato, differently from speakers to convey emotional information~\cite{speech-vs-singing}. Such characteristics provide interpretable descriptors of vocal expression that are specific to singing and are not necessarily captured by generic acoustic representations. Consequently, an interpretable SiER system should explicitly incorporate singing-specific features alongside more general acoustic descriptors.

Some studies have pursued interpretable, feature-based SiER. Specifically, \cite{acoustic_feats_singing} extract a set of openSMILE~\cite{eyben2010opensmile} descriptors from controlled opera recordings, spanning loudness, spectral flux and entropy, band energies, alpha ratio, Hammarberg index, and others, ranking them by importance and evaluating an \textsc{SVM} classifier under a leave-one-singer-out protocol. Another study extracts timbre-related spectral descriptors from \textsl{RAVDESS} singing samples and uses \textsc{SHAP} with an \textsc{XGBoost} classifier to derive feature contributions~\cite{singing_shap}. Our analysis corroborates several of their selected descriptors, yet neither study explicitly takes into account vocal technique. 

\begin{figure*}[!t]
    \centering
    \includegraphics[width=0.80\linewidth]{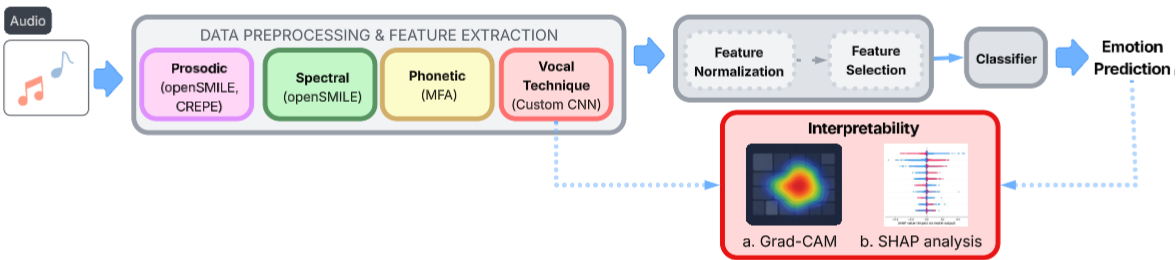}
    \caption{Overview of the proposed framework.}
    \label{fig:methodology}
\end{figure*}

In this work, we propose an interpretable framework for emotion recognition in singing based on explicitly measurable musical and singing-specific acoustic features (Figure~\ref{fig:methodology}). The framework extracts interpretable descriptors of the singing signal and uses them for emotion classification, enabling predictions to be examined in terms of identifiable characteristics of vocal expression rather than solely learned latent representations. In particular, we investigate whether incorporating domain-specific features can provide an effective and interpretable representation for SiER. Evaluations across diverse singing datasets indicate that the proposed features improve cross-singer stability, and 
reveal associations between emotional expression and vocal execution, with technique descriptors serving as an interpretable vocabulary for explaining expressivity in singing. 

\section{Methodology}
Our proposed framework consists of three main stages. First, we extract a broad set of descriptors correlated with the singing voice, spanning four perceptual dimensions. Secondly, we apply a feature selection process to obtain a compact feature set that supports interpretability without sacrificing predictive power. Finally, we employ explainable AI techniques to interpret the model predictions and assess the learned vocal technique representations.\looseness-1

\subsection{Feature extraction}

We extract 130 scalar features from audio resampled to 22.05~kHz, organized into four categories: prosodic, spectral, phonetic, and vocal technique features. For corpora lacking phoneme or technique annotations, these are estimated with the Montreal Forced Aligner (MFA)~\cite{mfa} and a dual-input CNN (Section~\ref{subsec:model}), respectively. All features are $z$-score normalized per singer, which reduces singer-specific variability while preserving within-singer differences and improves cross-corpus robustness~\cite{cross_corpus_strategies}.\looseness-1

Prosodic features describe pitch, intonation, and intensity, which shape perceived expressiveness in singing~\cite{SUNDBERG2013}. Using the \textsc{openSMILE eGeMAPS} configuration~\cite{egemaps}, \textsc{librosa}~\cite{mcfee2015librosa} and \textsc{CREPE}~\cite{crepe}, we compute features such as $f_0$ statistics,  pitch-change rate ($\Delta f_0$), jitter and shimmer. 
Spectral features characterize timbre, which varies with vocal effort and singing technique~\cite{acoustics_sund}. Using \textsc{openSMILE} and \textsc{librosa}, we compute the alpha ratio, formant frequencies, spectral flux and zero-crossing-rate statistics. Phonetic features, derived from phoneme boundaries, describe articulation and timing, and include the mean and standard deviation of phoneme and pause durations and the articulation rate. Vocal technique features capture expressive devices such as vibrato~\cite{vibrato_sund}. From the CNN's frame-level probabilities across six techniques (vibrato, glissando, mix, pharyngeal, falsetto, and breathy), we compute four statistics (mean, standard deviation, maximum, and peak factor). We also compute overall technique entropy by aggregating the presence ($\hat{y}_{t, c}$) of each technique $c \in \mathcal{C}$ across all $T$ frames and normalizing by the overall technique sum:
\begin{equation}
    S_c = \sum_{t=1}^{T} \hat{y}_{t, c}\,, \qquad p_c = \frac{S_c}{\sum_{k \in \mathcal{C}} S_k}
\end{equation}
The recording-level entropy is then calculated via Shannon entropy:
\begin{equation}
    H = -\sum_{c \in \mathcal{C}} p_c \ln(p_c)
\end{equation}
and quantifies the diversity of vocal technique presence. The complete feature list is available in our GitHub repository~\footnote{\href{https://github.com/FotiniKrp/Interpretable_Singing_Emotion_Recognition}{\nolinkurl{https://github.com/FotiniKrp/Interpretable_Singing_Emotion_Recognition}}}.

\subsection{Feature selection and model interpretability} \label{subsec:featsel}

Feature selection uses a nested Leave-One-Singer-Out (\textsc{LOSO}) framework to identify features that are consistently useful across singers, rather than ones that happen to fit a particular singer split. In the inner loop specifically, tree-based classifiers compute feature importance scores, and a frequency counter tracks how often each descriptor ranks among the top-$N$ candidates across inner splits, retaining only those with high cross-singer stability. Finally, across all outer singer iterations, we aggregate fold selections to retain the most stable descriptors. We apply this procedure both per feature category ($N = 10$) and to the complete feature set ($N = 40$). To reduce redundancy and improve interpretability, highly correlated features were pruned using pairwise Pearson correlation analysis (threshold = 0.9). The resulting feature set is then used to train a classifier for the final emotion prediction. This compact set supports a clearer interpretation of model predictions, in particular with the help of \textsc{SHAP}~\cite{shap} analysis to identify which features contribute to the classifier's predictions. We also employ \textsc{Grad-CAM}~\cite{grad-cam} to examine whether the vocal technique extractor relies on plausible acoustic cues rather than spurious patterns.
        
\section{Experimental Setup}    
\subsection{Dataset selection}\label{subsec:datasets}

To investigate singing emotion across diverse conditions and styles, we employ three datasets. As a baseline, we use \textsl{RAVDESS}~\cite{ravdess}, specifically its singing subset, comprising 1,012 English clips from 23 actors portraying six discrete emotions through acted expression. To complement these recordings with more realistic and diverse singing conditions, we additionally employ \textsl{GTSinger}~\cite{gtsinger}. For emotion classification, we selected the English, French, and German subsets, which are Western European languages, allowing a controlled evaluation. This subset encompasses modern pop and bel canto singing from seven vocalists (3 tenor, 2 alto, 2 soprano) and contains 1,228 performances (segmented into 10,656 clips), each labeled with a global emotion (happy or sad). To train the vocal technique extractor, we utilize the Chinese, English, Japanese, and French subsets, which offer a more balanced and diverse representation of vocal technique. Finally, we hold out its Italian subset to evaluate the CNN's ability to generalize to an unseen language. Lastly, \textsl{VocalSet}~\cite{wilkins2018vocalset} serves as an out-of-domain evaluation set to validate and analyze the learned vocal technique representations. Specifically, we utilize the 2,553 vocalizations across 20 singers from the filtered \textsl{Annotated-VocalSet} subset~\cite{vocalset_annotated}.

\subsection{Models}\label{subsec:model}
The vocal technique extractor is trained on 110,000 non-overlapping segments from \textsl{GTSinger}, using a 1.5~s window to provide sufficient temporal context for techniques such as vibrato while remaining a practical training unit. We use an 85:15 train/val split and the Adam optimizer with learning rate $10^{-4}$. To address the pronounced class imbalance, we employ Focal Loss with $\alpha=0.25$ and $\gamma=2.0$~\cite{focal_loss}:
\begin{equation}
    \mathcal{L}_{\text{FL}}(p_t) = -\alpha (1 - p_t)^\gamma \log(p_t)
\end{equation}
where $p_t$ is the sigmoid probability of the correct label for each technique. 
To further mitigate the effects of class imbalance, we employ early stopping on validation Macro-F1. Our CNN combines Mel-spectrogram and pitch contour representations through a dual-input architecture, inspired by prior work~\cite{primadnn}. The two streams are integrated through late fusion followed by a sigmoid output layer for multi-label classification, as multiple techniques can co-occur. For the emotion classification we employ \textsc{XGBoost} for \textsl{RAVDESS} (100 estimators, depth 6, learning rate 0.1, subsample/colsample = 0.8; feature-selection variant: depth 4, no stochastic sampling) and \textsc{Random Forest} for \textsl{GTSinger} (200 estimators, depth 12, min. leaf samples 10, balanced class weights; feature-selection variant: 50 estimators, depth 6, default min. leaf samples). 

We evaluate the models using a \textsc{LOSO} protocol \cite{cross_corpus_strategies}, where each singer is held out for testing while the remaining singers are used for training. Performance is measured using Unweighted Average Recall (UAR) and Macro-F1, and reported as the mean and standard deviation across all singer folds. Confidence intervals are estimated via singer-level bootstrapping  (10,000 iterations). For \textsl{GTSinger}, clip predictions are aggregated using majority voting to obtain the performance prediction before computing evaluation metrics. 
%To further test generalizability, the selected feature sets are also evaluated cross-corpus.\looseness-1

\section{Results}
\subsection{Performance}\label{subsec:performance}

On \textsl{RAVDESS}, we compare the predictive power of two compact subsets derived via our nested \textsc{LOSO} feature selection: a speech-optimized set, selected from the eGeMAPS set, and a singing-optimized set, selected from our proposed feature set. As shown in Table~\ref{tab:feature_set_comparison}, the singing-optimized set achieves a higher UAR (79.6\% with 95\% CI: 76.9--82.2\%) than the speech-optimized set (76.2\% with 95\% CI:  72.2--80.0\%), while also exhibiting lower cross-singer variability (singing-to-speech SD ratio = 0.68 with 95\% CI: 0.452--0.927). An ablation study (Table \ref{tab:ablation}) shows that prosody dominates baseline accuracy, spectral and phonetic features provide complementary cues, while technique descriptors slightly reduce cross-singer deviation.

\begin{table}[!h]
\centering
\caption{Feature Set Performance on \textsl{RAVDESS} and \textsl{GTSinger}.}
\label{tab:feature_set_comparison}
\footnotesize
\renewcommand{\arraystretch}{1.15}
\setlength{\tabcolsep}{5.5pt}
\begin{tabular}{@{}llcccc@{}}
\toprule
\textbf{Dataset} & \textbf{Feature Set} & \textbf{\#} & \textbf{UAR (\%)} & \textbf{SD (\%)} & \textbf{Macro-F1} \\
\midrule
\multirow{2}{*}{\textsl{RAVDESS}} 
 & Speech-Opt.   & 19  & 76.2 & 9.5 &0.75 \\
 & \textbf{Singing-Opt.}  & 18  & \textbf{79.6} & \textbf{6.5} & \textbf{0.79}
 \\
 % & Full Set      & 130 & \textbf{82.4} & 8.2 & \textbf{0.82} \\
 % & eGeMAPS   & 88 & \textbf{77.8} & 8.8 & \textbf{0.77} \\
\midrule
\multirow{2}{*}{\textsl{GTSinger}} 
 & Speech-Opt.   & 20  & 60.2 & 5.8 & 0.56 \\
 & \textbf{Singing-Opt.} & 19  & \textbf{61.0} & \textbf{3.3} & \textbf{0.60 }\\
\bottomrule
\end{tabular}
\end{table}

\begin{table}[!h]
\centering
\caption{Ablation Study on the Singing-Optimized Feature Set.}
\label{tab:ablation}
\footnotesize
\renewcommand{\arraystretch}{1.15}
\setlength{\tabcolsep}{5.5pt}
\begin{tabular}{@{}llcc@{}}
\toprule
\textbf{Type} & \textbf{Feature Set} & \textbf{UAR (\%)} & \textbf{SD (\%)} \\
\midrule
\multirow{4}{*}{Single}
 & Only Techniques  & 35.0 & \textbf{8.6} \\
 & \textbf{Only Prosody }    & \textbf{66.4} & 11.1 \\
 & Only Phonetic    & 32.8 & 9.9 \\
 & Only Spectral    & 53.5 & 11.0 \\
\midrule
\multirow{4}{*}{All-but-1}
 &\textbf{All exc. Techniques}& \textbf{79.4} & 7.0 \\
 & All exc. Prosody    & 62.5 & 7.6 \\
 & All exc. Spectral   & 73.2 & 10.0 \\
 & All exc. Phonetic   & 76.8 & \textbf{5.5} \\
\bottomrule
\end{tabular}
\end{table}

%A per-class analysis reveals that the singing-optimized set improves the F1-scores of calm (+0.05), sad (+0.05), and neutral emotions (+0.01), at a small cost to high-arousal classes (angry $-$0.03, happy $-$0.02, fearful $-$0.02). 
%To quantify the predictive contribution of each feature category, we performed an ablation study on the singing-optimized feature set. Table~\ref{tab:ablation} shows that prosodic and spectral descriptors dominate, while technique descriptors are comparatively weaker. This gap may partly reflect \textsl{RAVDESS}'s limited technique diversity. We therefore revisit the role of vocal technique descriptors on \textsl{VocalSet} and \textsl{GTSinger} in the following sections.
We replicate the feature selection procedure on GTSinger. As shown in Table~\ref{tab:feature_set_comparison}, while the speech-optimized feature set attains a comparable UAR to the singing-optimized set, it exhibits higher cross-singer variance and a lower macro-F1 score. %highest mean UAR but also the highest variability, while the \textsl{RAVDESS} singing-optimized set is less accurate but markedly more stable. The \textsl{GTSinger}-native set falls between the two, reflecting a more balanced trade-off. Notably, the speech-optimized set shows a large style-dependent gap (63.0\% bel canto vs. 58.2\% pop, dropping to near-chance 52.6\% on the most difficult pop singer), whereas the singing-optimized set remains comparatively stable across styles (60.9\% vs. 61.2\%), though these subgroups are limited in size ($n=3$ bel canto, $n=4$ pop).
Notably, the speech-optimized set shows a style-dependent gap (63.0\% bel canto vs. 58.2\% pop, dropping to near-chance 52.6\% on the most difficult pop singer), whereas the singing-optimized set remains comparatively stable across styles (60.9\% vs. 61.2\%), though these subgroups are limited in size ($n=3$ bel canto, $n=4$ pop).

\subsection{Interpretability analysis}\label{subsec:analysis}

\begin{figure}[!h]
  \centering
  \begin{minipage}[b]{0.49\linewidth}
    \centering
    \includegraphics[width=\linewidth]{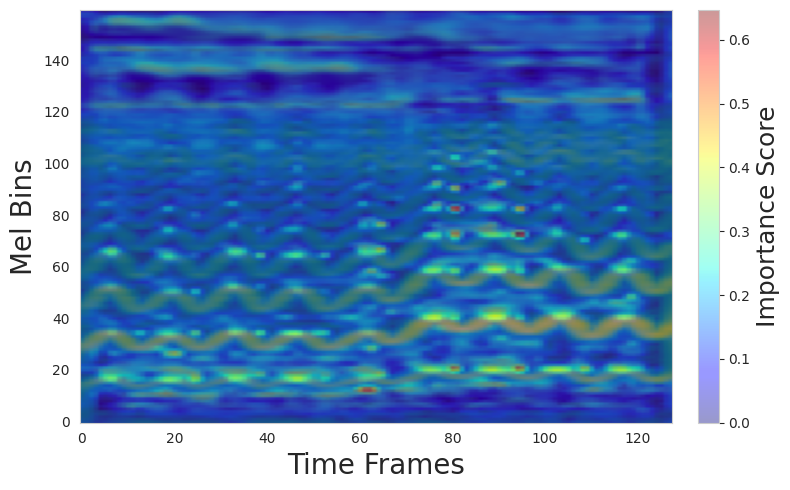}
  \end{minipage}
  \hfill
  \begin{minipage}[b]{0.49\linewidth}
    \centering
    \includegraphics[width=\linewidth]{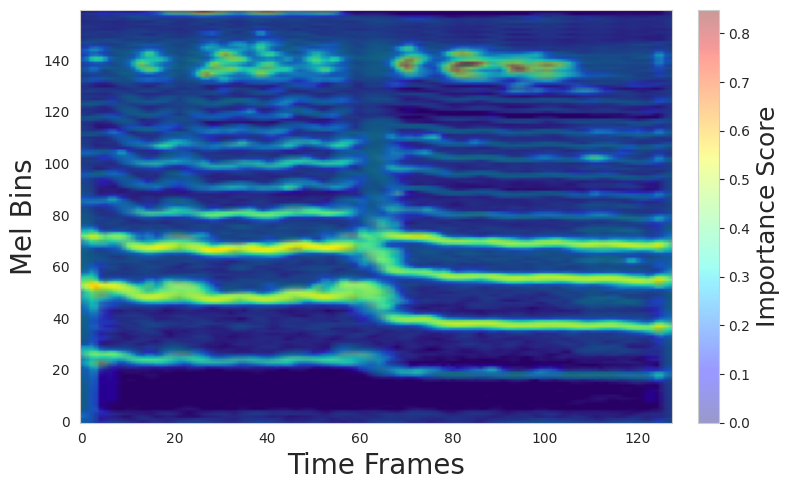}
  \end{minipage}
   \caption{\small \textsc{Grad-CAM} on \textsl{VocalSet}: (a) vibrato, (b) belt.}
\label{fig:grad-cam-samples}
\end{figure}

\begin{figure}[!b]
\centering
\includegraphics[width=0.65\linewidth]{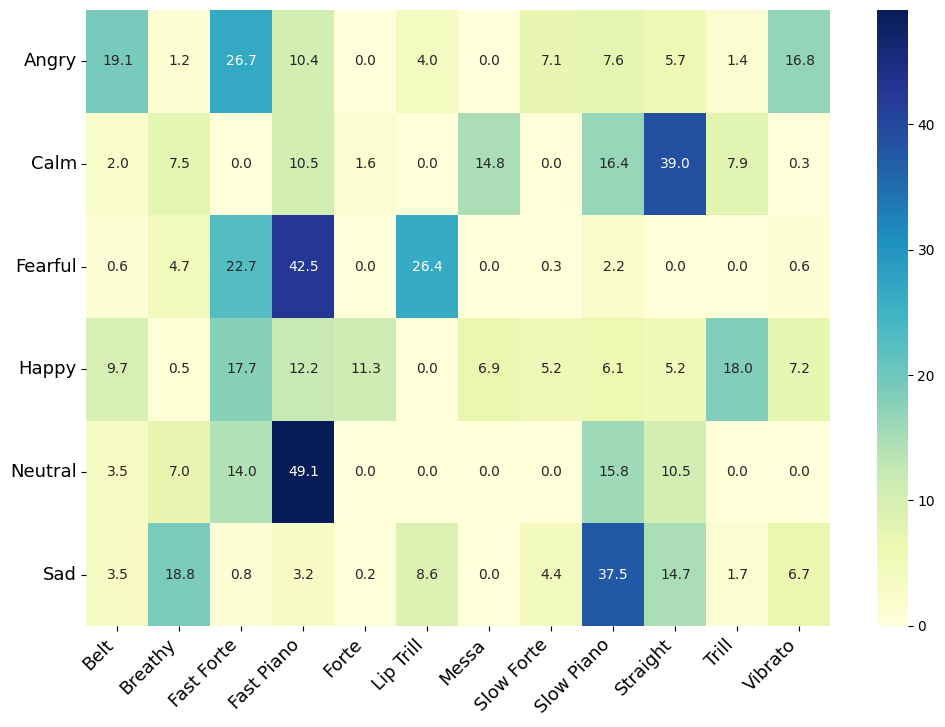}
\caption{\small Distribution of predicted emotion categories across \textsl{VocalSet} ground-truth techniques for the \textsl{RAVDESS} singing-optimized model.}
\label{fig:corr_vocalset}
 \end{figure}

When evaluated on the held-out Italian subset of \textsl{GTSinger}, the vocal technique extractor achieved a Macro-F1 of 0.42 and a Macro-Recall of 0.56 across multi-label frame predictions. We employ \textsc{Grad-CAM} to visualize the time–frequency regions contributing most strongly to the CNN's predictions, as a qualitative check that the model attends to acoustically meaningful regions. Figure~\ref{fig:grad-cam-samples} shows representative \textsl{VocalSet} samples. For the vibrato example, the \textsc{Grad-CAM} map highlights the peaks of the periodic frequency modulations associated with the vibrato pattern (the least-represented class during training). For the belt sample, the \textsc{Grad-CAM} computed with respect to the pharyngeal class shows two distinct activation patterns: continuous horizontal bands in the mid-frequency mel bins, consistent with pressed phonation  \cite{grad-cam_phonation}, and a localized cluster of high- activation in the highest mel bins, which may instead reflect flow phonation, commonly associated with belting technique. We further evaluate the representations learned from \textsl{RAVDESS} by applying both the speech and singing-optimized models to \textsl{VocalSet} and examining the associations between predicted emotions and ground-truth vocal techniques. As shown in Figure \ref{fig:corr_vocalset}, high-arousal emotions such as anger are more strongly mapped to dynamic vocal techniques, including belt, fast articulated forte, and vibrato, whereas low-arousal emotions such as sadness are associated with more subdued techniques, including breathy, slow legato piano, and straight tone. We additionally compare the mappings of the speech- and singing-optimized models using mutual information (MI). The singing-optimized model achieves higher overall MI with the technique labels (0.474 vs. 0.352), suggesting that the extracted singing-specific descriptors capture genuine technique-related information. 

\begin{figure}[!h]
  \centering
  \begin{minipage}[b]{0.49\linewidth}
    \centering
    \includegraphics[width=\linewidth]{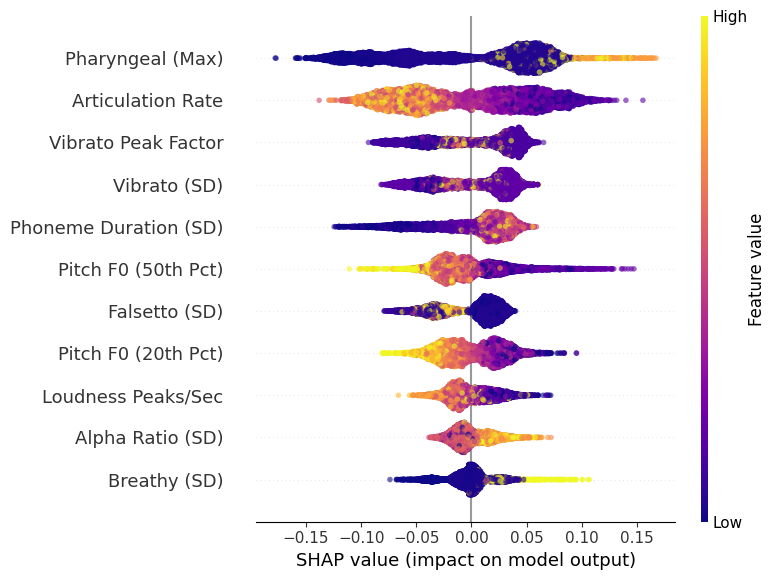}
  \end{minipage}
  \hfill
  \begin{minipage}[b]{0.49\linewidth}
    \centering
    \includegraphics[width=\linewidth]{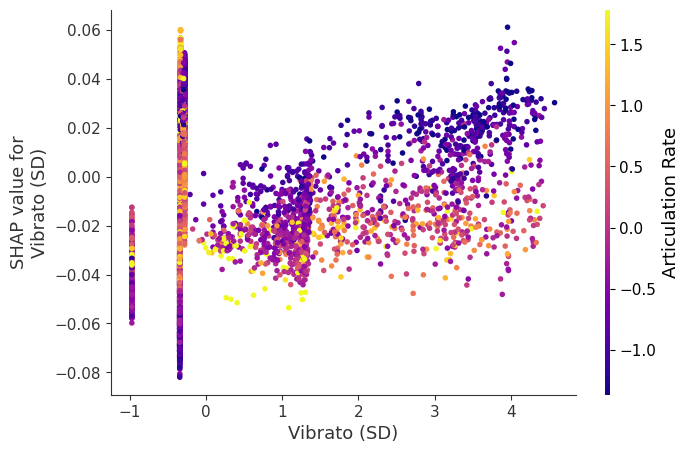}
  \end{minipage}
   \caption{ \textsc{SHAP} for the Singing-Opt. Feature Set: (a) beeswarm for the sad class (b) vibrato dependence for sad.}
\label{fig:shap_combined}
\end{figure}

For \textsl{GTSinger}, we train a classifier, with each clip inheriting the label of its parent performance. \textsc{SHAP} analysis 
(Fig. ~\ref{fig:shap_combined}a) shows that predictions for happy are associated with higher pitch levels, elevated articulation rate, more frequent loudness peaks, and lower variability in phoneme duration, indicating a more stable and energetic expression. Conversely, sad predictions exhibit a more nuanced profile characterized by a higher pharyngeal quality, greater variability in breathiness, lower pitch and a reduced articulation rate. 
Utilizing a \textsc{SHAP} dependence plot (Fig.~\ref{fig:shap_combined}b), we observe that the attribution of vibrato variability is mediated by articulation rate: at slower rates, elevated vibrato variability is associated with sad predictions (consistent with a more plaintive delivery), whereas at elevated rates it drives happy predictions, consistent with a more vibrant expression.

\subsection{Discussion}
Vocal technique descriptors contribute to our framework's interpretability. Although they provide limited numerical gains in raw accuracy in GTSinger (Section~\ref{subsec:performance}), they establish acoustically coherent mappings between predicted emotion and ground-truth technique on \textsl{VocalSet} and reveal meaningful acoustic patterns (Section~\ref{subsec:analysis}). Their apparent association with low-valence emotional states suggests that they offer complementary information beyond conventional prosodic cues. In particular, pharyngeal technique emerges as the top predictor in our SHAP analysis, consistent with prior work~\cite{resonance_emotion} highlighting resonance manipulation as an important aspect of expressivity in singing. While~\cite{vibrato} demonstrates that vibrato parameters vary with emotional engagement, our analysis further suggests that vibrato's relationship with predicted emotion depends on articulation rate. Other consistently selected descriptors, including pitch, timbre, intensity, and articulation rate, align with acoustic cues previously linked to emotional expression in vocal communication and music performance~\cite{speechvsmusic}. Cross-corpus comparisons also indicate that feature relevance varies with singing style, as reflected in the descriptor types favored by each corpus (Table \ref{tab:feature_sets}) and the lower cross-style stability of the speech-optimized feature set (Section~\ref{subsec:performance}). However, disentangling stylistic differences from other corpus-related factors requires further investigation.~\looseness-1

\begin{table}[!h]
\centering
\caption{Representative Descriptors of the Singing-Optimized Sets.}
\label{tab:feature_sets}
\footnotesize
\renewcommand{\arraystretch}{1.15}
\setlength{\tabcolsep}{4pt}
\begin{tabularx}{\columnwidth}{@{}l X@{}}
\toprule
\textbf{Subset} & \textbf{Selected Acoustic \& Technique Descriptors} \\
\midrule
Common (6) &
articulation rate, phoneme duration (SD), loudness peaks/s, local shimmer, alpha ratio ($\sigma_{\text{norm}}$), and pharyngeal (max). \\
\midrule
\textsl{GTSinger} (13) &
vibrato (SD, peak), falsetto (SD), breathy (SD), $f_0$ percentiles (20th, 50th), voiced segments/s, local jitter, and $\sigma_{\text{norm}}$ of HNR, spectral slope (0--500\,Hz), and harmonic differences ($H_1\text{--}H_2$, $H_1\text{--}A_3$). \\
\midrule
\textsl{RAVDESS} (12) &
$f_0$ ($\sigma_{\text{norm}}$, 80th pct., mean deviation), mean HNR, spectral flux, alpha ratio, $F_3$ amplitude ($\sigma_{\text{norm}}$), breathy (mean), falsetto/glissando (max), and technique entropy. \\
\bottomrule
\end{tabularx}
\end{table}

Several limitations warrant consideration. The observed relationships represent statistical associations rather than causal mechanisms, and generalizability is constrained by the limited number of singers, the acted emotions in \textsl{RAVDESS}, and the restricted emotion categories in \textsl{GTSinger}. Moreover, vocal technique features rely on CNN-inferred probabilities, potentially introducing label noise.\looseness-1

\section{Conclusion}

We presented an interpretable framework for singing emotion recognition using acoustic and vocal technique descriptors. A compact, singing-specific feature set provided a useful representation for emotion recognition while reducing variability across singers. Cross-corpus analyses revealed associations between predicted emotions, vocal techniques, and acoustic characteristics, illustrating the potential of interpretable descriptors to uncover meaningful patterns in vocal expression. Future work should evaluate the framework across more diverse languages, musical genres, and vocal traditions.

\bibliographystyle{IEEEtran}
\bibliography{references}
\end{document}